# A Web-Based Resource Model for eScience: Object Reuse & Exchange

Carl Lagoze
Information Science
Cornell University
lagoze@cs.cornell.edu

Herbert Van de Sompel
Research Library
Los Alamos National Lab
herbertv@lanl.gov

Michael Nelson
Computer Science
Old Dominion University
mln@cs.odu.edu

Simeon Warner
Information Science
Cornell University
simeon@cs.cornell.edu

Robert Sanderson
Department of Computer Science
University of Liverpool
azaroth@liv.ac.uk

Pete Johnston
Eduserv Foundation
pete.johnston@eduserv.org.uk

Despite the high "hype to reality" ratio, Web 2.0 [37] represents a fundamental change in the Web. Originally conceived as a technology for linking together documents [8], the Web has morphed into a rich network of distributed services, data, and semantic relationships, glued together via social participation [35]. Social applications such as blogs and wikis and social sites such as facebook and Flickr demonstrate how the Web has evolved into a socio-technical network [9], a context where *object centered sociality* [16] takes place and is subsequently recorded.

The power of social participation in Web 2.0 has had substantial impact on cyberinfrastructure applications [15, 34] that until recently leveraged the Web as a mainly technical platform. For example, digital library practitioners, who initially used the Web mainly as a technology for the deployment of traditional concepts (documents, repositories, catalogs, metadata, etc.) [10], now recognize a new information paradigm in which social interaction and the information context that it imprints are as important as collections of artifacts [22, 31]. Similarly, scholarly communication, a phenomenon that has long been described in socio-technical terms [20], is changing in response to Web 2.0. Traditional read-only journal and conference papers, and peer review are being supplemented and even replaced by online mechanisms for contribution, participation, and feedback [11, 18]. Finally, the grids of eScience are being reformulated to include a Web 2.0 inspired "Architecture of participation that encourages user contribution" [17].

We argue that the increasing synergy between Web 2.0 and cyberinfrastructure must be reflected in common data models, protocols, and standards. This will allow the artifacts of eScience (data, documents, tools) to be exposed to the broad Web audience via widely deployed mechanisms such as Atom feeds and mashups, which make them accessible outside their original context – for example, for teaching and learning [21]. The reverse is also true. It must be easy to include resources from the general web into eScience. Failure to integrate eScience with the mainstream Web will serve to isolate it to the so-called "invisible web" [32, 39], where it will be undiscoverable by mainstream search engines. As a result, the goal of "removing access barriers" promoted by open access initiatives [1] will not be accomplished.

### *Identifying and Describing Compound Objects*

Our work in Open Archives Initiative – Object Reuse and Exchange (OAI-ORE) - focuses on one particular aspect of this shared infrastructure. This is the specification of a data model and a suite of implementation standards to identify and describe compound objects – objects that aggregate multiple sources of content including text, images, data, visualization tools, and the like. We argue elsewhere [7, 41-43], as have others in the eScience community [33, 44], that such *aggregations* are an essential product of eScience. Furthermore, while the notion of an *aggregation* is not explicit in the Web Architecture, it is prevalent across general Web space. For example, a "photo" in Flickr is an aggregation of multiple renditions in different sizes, and that photo is aggregated along with other "photos" into a "collection". Similarly, the blog entry that we think of as a singleton is in fact an aggregation composed of the original entry combined with multiple comments (and comments on comments). That blog entry is itself aggregated in a subject partition of a blog. Thus, a suite of standards regarding aggregations will benefit both the eScience and Web community.

We note that despite their logical presence on the Web, these aggregations are not recognized in the Web architecture [19], which defines the following notions: Resource, an item of interest; URI, a global identifier for a Resource, commonly using the HTTP scheme; Representation, a datastream corresponding to the state of a Resource at the time its URI is dereferenced; and Link, a directed connection between two Resources, which, when extended with types, forms the foundation of the Semantic Web.

The OAI-ORE specifications endow aggregations with two attributes that we consider essential to their utility in the shared eScience/Web context.

- *Identity:* Identity is used in the scholarly context for expressing citation, lineage, and rights. As noted above, it is a core concept in the Web architecture for browser access, hyperlinking, and in the semantic web to express assertions, or semantic relationships, between resources. Despite the existence of many identity schemes in both the digital and physical information space (DOIs, Handles, ISBNs, etc.) [38], OAI-ORE specifies resolvable URIs to identify aggregations, thereby establishing an aggregation as a Web Resource that can be accessed and linked to like any URI-identified Resource.

- *Boundary:* The ability to deterministically enumerate the constituents of an aggregation is essential for eScience and related application areas. Boundary, for example, informs preservation services "what to preserve" and rights management applications "who is responsible for what". While not defined in the Web Architecture, the importance of boundary has also been acknowledged in the Web community. It is for example part of the requirement set of the Protocol for Web Description Resources (POWDER) [5] work, which aims to provide mechanisms to publish properties shared by a set of Web resources.

### *OAI-ORE Standards and Specifications*

Space restrictions prohibit a full description of the ORE specifications, which address these issues. The interested reader is referred to the full set of ORE documents at [25]. We briefly summarize the content of these documents.

*Data Model* - The ORE Data Model [23] makes it possible to associate an identity with aggregations of Web resources and to describe their structure and semantics. It does this by introducing the *Resource Map* (ReM), which is a resource identified by a URI (say ReM-1) that encapsulates a set of RDF statements [30]. The notion of associating a URI with a set of RDF statements is based on the concept of a *named graph* developed in the Semantic Web community [12]. The creation of a Resource Map instantiates an aggregation as a resource with a URI distinct from the Resource Map, enumerates the constituents of the aggregation, and defines the relationships among those constituents. Although a Resource Map may assert a variety of RDF statements, it must assert a set of statements that define the *aggregation graph.* These statement define a sub-graph relating the Resource Map to the Aggregation via the `ore:describes` predicate, the Aggregation to its constituent Aggregated Resources via the `ore:aggregates` predicate, and associate the Resource Map with key metadata properties: `dcterms:creator` and `dcterms:modified`.

*Serialization* – Serialization provides a means of transmitting, introspecting upon, and storing ORE data model-based descriptions of aggregations. Because the ORE data model is based on RDF triples, the RDF/XML [6] expression of these triples is a natural and fully expressive serialization [29]. In addition, to make Resource Maps more widely accessible we define a somewhat less expressive serialization in the popular XML-based Atom syndication format [26, 28, 36].

*HTTP Implementation* – The ORE Data Model is not specific to any implementation protocol. However, the use of HTTP URIs to identify ORE Aggregations and Resource Maps leverages the extensive infrastructure and tools of the current World Wide Web. HTTP provides mechanisms that allow the Aggregation, which is a non-information resource in the sense of the Web Architecture, to yield or redirect to a Resource Map as required by the ORE Model. We describe different HTTP implementation scenarios that differ in the server requirements needed to support them, and in the URI structure that results [24].

*Discovery* - Discovery is a precondition for the use of Resource Maps. There is no single, best method for discovering Resource Maps, and we expect best practices for discovery to evolve over time. The Resource Map Discovery Document [27] covers a variety of suggested Resource Map discovery mechanisms, grouped into the categories of Batch Discovery, Resource Embedding and Response Embedding.

### *Deployment, Experimentation, and Implementation*

At the time of writing this paper the OAI-ORE specifications are still in beta development stage. Full production release is scheduled for end of September 2008.

Because of this pre-production state there have yet to be a large number of full implementations, since members of the various communities that have shown active interest in OAI-ORE are awaiting final release. There have been however a number of experimental implementations, as follows:

- Enhancing the Zotero citation manager [13] to allow the user to download any number of constituent resources of the compound object.
- Archiving compound information objects as they evolve over time [40].
- Constructing a compound eScience object and publishing it to institutional repositories [14].
- Demonstrating the utility of ORE for thesis description, submission, and publication (TheOREM) [2].
- Creating Resource Map descriptions of JTSOR's holdings and ingesting them into the DSpace institutional repository system via the SWORD protocol [4].
- Demonstrating OAI-ORE powered functionality at the end-user level (e.g. in a browser) [3].

In addition, several larger scale applications are planned. The OREChem Project funded by Microsoft and involving Cambridge, Cornell, Indiana, Penn State, and Southampton is planning to use ORE as the basis of interoperability among variety of chemistry molecule repositories, and will build innovative applications on that infrastructure. And, OAI-ORE is a key infrastructure component of finalists for the DataNet program of the National Science Foundation.

Clearly, the success of the OAI-ORE work remains to be proven, in particular with full implementations in the eScience and general cyberinfrastructure community. However, the work has the attention of a spectrum of communities and organizations. Feedback throughout a lengthy alpha and beta testing program has been generally positive and effective at evolving the work to meet the real needs of the deployment community.

### *Acknowledgments*

OAI-ORE is supported by the Andrew W. Mellon Foundation, the Coalition for Networked Information, Microsoft, and the National Science Foundation (IIS-0430906). The authors acknowledge the contributions to the OAI-ORE effort from the ORE Technical Committee, Liaison Group and Advisory Committee.